\documentstyle[12pt,epsfig,equation]{article}
\def \LSP{\widetilde{\chi}_1^0}
\def \N2{\widetilde{\chi}_2^0}
\def \CH{\widetilde{\chi}^{\pm}}
\def \CH1{\widetilde{\chi}_1^{\pm}}   
\def \SNU{\tilde{\nu}}

\def \LR{\tilde{l}_R}
\def \LL{\tilde{l}_L}

\def \UR{\tilde{u}_R}
\def \UL{\tilde{u}_L}
\def \DR{\tilde{d}_R}
\def \DL{\tilde{d}_L}
\def \BR{\tilde{b}_R}
\def \BL{\tilde{b}_L}
\def \GLU{\tilde{g}}
\def \MLSP{m_{\widetilde{\chi}_1^0}}
\def \MN2{m_{\widetilde{\chi}_2^0}}
\def \MCH{m_{\widetilde{\chi}^{\pm}}}
\def \MCH1{M_{\widetilde{\chi}_1^{\pm}}}
\def \MSNU{m_{\tilde{\nu}}}
\def \MSELR{m_{\tilde{e}_R}}
\def \MSELL{m_{\tilde{e}_L}}
\def \MLL{m_{\tilde{l}_L}}

\def \MGLU{m_{\tilde{g}}}
\def \MSQ{m_{\tilde{q}}}

\def \MUR{m_{\tilde{u}_R}}
\def \MUL{m_{\tilde{u}_L}}
\def \MDR{m_{\tilde{d}_R}}
\def \MDL{m_{\tilde{d}_L}}

\def \MBL{m_{\tilde{b}_L}}
\def\lsim{\:\raisebox{-0.5ex}{$\stackrel{\textstyle<}{\sim}$}\:}
\def\gsim{\:\raisebox{-0.5ex}{$\stackrel{\textstyle>}{\sim}$}\:}
 
\def \M0{m_0}
\def \MG{m_{1/2}}
\def \MGUT{M_G}

\def \sio{\sigma_0}

\def \S3l{\sigma_{3l}}

\newcommand{\be}{\begin{equation}}
\newcommand{\ben}{\begin{subequations}}
\newcommand{\een}{\end{subequations}}
\newcommand{\beq}{\begin{eqalignno}}
\newcommand{\eeq}{\end{eqalignno}}
\newcommand{\ee}{\end{equation}}

\begin{document}
\setcounter{page}{0}
\thispagestyle{empty}

\begin{flushright}
IFT--P058/99 \\
TIFR/TH/99--35 \\
MRI--PHY/P990721 \\
July 1999\\
\end{flushright}
\vspace*{10mm}

\begin{center}
{\Large\bf Effects of $SO(10) \ D-$Terms on SUSY Signals at the Tevatron}\\
\vspace*{10mm}
Amitava Datta$^1${\footnote {\tt adatta@juphys.ernet.in}}, 
Aseshkrishna Datta$^2${\footnote {\tt asesh@mri.ernet.in}},
Manuel Drees$^3${\footnote {\tt drees@ift.unesp.br} }, and
D.P. Roy$^4${\footnote {\tt dproy@theory.tifr.res.in}} \\
\hspace*{5mm}
$^1${\it Department of Physics, Jadavpur University, Calcutta 700 032,
India}\\
$^2${\it Mehta Research Institute, Chhatnag Road, Jhusi, Allahabad 211 019,
India}\\
$^3${\it IFT, Univ. Estadual Paulista, 01405--900 S\~ao Paulo, Brazil} \\
$^4${\it Tata Institute of Fundamental Research, Mumbai 400005, India}

\end{center}
\hspace*{15mm}

\begin{abstract}
We study signals for the production of superparticles at the Tevatron
in supergravity scenarios based on the Grand Unified group $SO(10)$.
The breaking of this group introduces extra contributions to the
masses of all scalars, described by a single new parameter. We find
that varying this parameter can considerably change the size of
various expected signals studied in the literature, with different
numbers of jets and/or charged leptons in the final state. The ratios
of these signal can thus serve as a diagnostic to detect or constrain
deviations from the much--studied scenario where all scalar masses are
universal at the GUT scale. Moreover, under favorable circumstances
some of these signals, and/or new signals involving hard $b-$jets,
should be observable at the next run of the Tevatron collider even if
the average scalar mass lies well above the gluino mass.

\end{abstract}

\newpage
\section {Introduction}
\setcounter{footnote}{0}

The search for Supersymmetry (SUSY) \cite{1}, the most attractive
candidate for physics beyond the Standard Model (SM), has top priority
in present day high energy experiments. From the non--observation of
sparticles at various colliders only lower limits on their masses have
been obtained so far.

{}From the model building point of view the limit $\MCH1 > 91$ GeV
\cite{2} on the mass of the lighter chargino $\CH1$ from LEP2 is of
special importance. This limit, in conjunction with the popular
assumption of gaugino mass unification \cite{1} with a common
gaugino mass $\MG$ at the GUT scale ($\MGUT$), implies that the gluino
($\GLU$) must be rather heavy with $\MGLU \gsim 300$ GeV.

The limit on $\MCH1$ becomes much weaker if the $\CH1$ happens to be
approximately degenerate with either the lightest supersymmetric
particle (LSP) assumed to be the lightest neutralino ($\LSP$), or with
a sneutrino ($\SNU$). The first possibility is strongly disfavored in
models with gaugino mass unification and radiative electroweak symmetry
breaking (see below). The second possibility occurs, e.g., in mSUGRA
models \cite{1} for small values of the scalar soft breaking mass
$\M0$. However, a recent result from the D\O\ collaboration \cite{3}
ruled out $\MGLU <$ 300 GeV for small $\M0$ from direct squark and
gluino searches in the multi--jet plus missing $E_T$ channel.

Like most other limits on squark and gluino production at hadron
colliders, this bound holds if the squarks are degenerate with each
other at $\MGUT$ with a common mass $\M0$. They are then also
approximately degenerate at the weak scale with average mass $\MSQ$.
Under the same assumption, for much heavier squarks ($\MSQ \gg \MGLU$)
only $\MGLU \lsim 200$ GeV can be excluded from present Tevatron
data. In this case, even gluino searches after the main injector
upgrade of the Tevatron (referred to hereafter as Tevatron Run II),
with an estimated integrated luminosity of 2 fb$^{-1}$ at $\sqrt{s}$ =
2 TeV, are not likely to improve the indirect limit on the gluino mass
from LEP2 searches, if gaugino mass unification holds. A further
luminosity upgrade, amounting to an integrated luminosity $\sim 25$
fb$^{-1}$ at $\sqrt{s} = $ 2 TeV (hereafter referred to as TeV33),
could reach somewhat higher, but not dramatically different, masses
through direct searches for gluinos and squarks \cite{4}.

As already mentioned, these estimated search limits have been derived
under certain assumptions, like gaugino mass unification and
degeneracy of squarks at $\MGUT$. Such assumptions about physics at high
scales, although attractive, may ultimately prove wrong. It is
therefore important to reexamine the search prospects at the Tevatron
and its various upgrades if some or all of these assumptions are
relaxed. If accompanied by suitable theoretical guidelines, this can
be accomplished while still avoiding an unmanageable proliferation in
the number of unknown parameters. Moreover, such an analysis indicates
the feasibility of obtaining glimpses of the SUSY breaking pattern at
a high scale using the Tevatron data. Finally, and most importantly,
such analyses are needed to assess the robustness of the SUSY search
strategies currently employed by Tevatron experiments.

One possible avenue would be to relax the gaugino mass unification
condition. This has been attempted by several groups \cite{5}.
However, at least in the framework of SUSY GUT's, the assumption of
gaugino mass unification appears quite natural, since it simply
follows if the GUT symmetry is respected by the SUSY breaking
mechanism at a high scale, irrespective of the specific choice of the
GUT group. In this paper we, therefore, assume that gaugino masses do
indeed unify.

The assumption that all soft breaking masses for the scalars have a
universal value $\M0$ at scale $\MGUT$ (referred to hereafter as the
conventional scenario), which in turn predicts approximately
degenerate squarks at the weak scale, is more model dependent. In the
framework of the popular $N = 1$ SUGRA GUT models \cite{6}, where SUSY
breaking in the hidden sector is transmitted to the observable sector
by gravitational--strength interactions alone, two assumptions are
required to obtain a degenerate scalar spectrum at scale
$\MGUT$. First, the K\"ahler metric has to be the same for all chiral
superfields. This ansatz drastically reduces the large number of
parameters that may otherwise be present in the Minimal Supersymmetric
Standard Model (MSSM), making the model much more predictive. It also
avoids problems with flavor changing neutral currents caused by
sparticle loops. This assumption guarantees degenerate sfermions at
the scale $\Lambda$ at which SUSY breaking is transmitted to the
visible sector. By taking sfermions to be degenerate at scale $\MGUT$
one therefore implicitly makes the second assumption that $\Lambda
\simeq \MGUT \simeq 2 \cdot 10^{16}$ GeV. However, from the
supergravity point of view, it is more natural to choose $\Lambda$ to
be near the Planck scale $M_P \simeq 2.4 \cdot 10^{18}$
GeV. Non-universality in scalar masses at scale $\MGUT$ can then arise
due to several reasons:

\noindent
i) The running of scalar masses between $M_P$ and $\MGUT$ \cite{7} may
lead to non-universality at $\MGUT$. Within the framework of an $SO(10)$
SUSY GUT, however, the resulting non-universality is expected to be
negligible for the first two generations of squarks and sleptons with
small Yukawa couplings. In contrast, the soft breaking masses of the
$SU(2)$ doublet Higgs bosons responsible for electroweak symmetry
breaking can be changed significantly, since they have to couple to
super-heavy GUT fields in order to generate very large masses for their
GUT partners (doublet--triplet splitting); moreover, these Higgs
bosons are usually assumed to reside in a different representation of
$SO(10)$ than the sfermions do, and hence are renormalized differently
by gaugino loop diagrams. The situation for third generation sfermions
is intermediate between these two extremes: they reside in the same
representation as first and second generation sfermions do, but at
least some of them have significant Yukawa couplings. Thus only the
first two generations of sleptons and squarks can safely be considered
to have a common mass at $\MGUT$, which we define to be $\M0$. The
soft breaking masses of third generation sfermions are likely to be
somewhat, but not very much, smaller, while those for the Higgs bosons
can be either larger or smaller than $\M0$. Fortunately, the physics
of the Higgs sector is relatively unimportant for direct squark and
gluino searches at Tevatron energies. In our study we will neglect the
(probably not very large) effect of the running between $\Lambda$ and
$\MGUT$ on third generation soft breaking terms.

ii) It has been known for some time that if the rank of a GUT group
(or some symmetry group broken at an intermediate scale) is reduced by
spontaneous symmetry breaking, one may obtain $D-$term contributions
to scalar masses \cite{8,9}. These will in general differ for
different members of the same GUT multiplet, leading to non-universal
squark and slepton masses at the symmetry breaking scale. The size of
these new contributions can be comparable to $\M0$. Note that these
nonuniversal terms are generation--independent, so that no additional
problems due to flavor changing neutral currents arise.

It has already been emphasized in \cite{10} that right--handed
down--type squarks ($\DR , ~\tilde{s}_R , ~\BR$, which are degenerate
in mass) may be considerably lighter than other species of squarks
($\UL, \DL$ or $\UR$) due to mechanism (ii). A specific model in which
this may happen is an $SO(10)$ SUSY GUT, breaking down to the SM either
directly or via the Pati--Salam group \cite{11} at an
intermediate scale \cite{9}. In addition, the L--type (left--handed)
or R--type (right--handed) sleptons may also turn out to be
considerably lighter, which may yield many novel experimental
signatures \cite{10,12}. The phenomenological consequences of $D-$terms
at high scales have also been studied in the first paper of \cite{5}.

In the present analysis we shall work within the framework of a
$SO(10)$ SUSY GUT breaking down to the SM directly. In this model
nonuniversality in the squark and slepton masses at a high scale can
be parametrized by only one extra parameter $D$ (compared to the
conventional scenario) \cite{9}. $SO(10)$ has several attractive
features: it allows to fit a complete generation of (s)quarks and
(s)leptons into a single representation of the group. Moreover, it
includes right--handed neutrinos, and thus naturally allows to
implement the see--saw mechanism \cite{see} to explain the small but
nonvanishing neutrino masses that seem to be needed to explain recent
data on neutrino oscillations. Direct breaking to $SU(3) \times SU(2)
\times U(1)$ is perhaps more natural, in view of the apparent
unification of all three gauge couplings in the MSSM in the absence of
any intermediate scale. Similar mass patterns are, however, expected
even if this GUT group first breaks to the Pati--Salam group.

Our main goal is to compare and contrast the SUSY signals predicted by
the nonuniversal scenario with those of the conventional one ($D = 0$)
which has 10 roughly degenerate squarks of L and R types excluding the
stop ($\tilde{t}$). We shall restrict our analysis to SUSY searches
at various phases of the Tevatron. Special attention will be paid to
the following issues: i) Is it possible to get a signal at Run II in
some region of the parameter space (not already excluded by LEP
searches) of the nonuniversal model with $\M0 \gg \MG$, which
corresponds to $\MSQ \gg \MGLU$ in the conventional scenario? ii) If
signals are seen at the Tevatron, can they always be accommodated in
both models, or might a distinction between the two be possible?

In the next section we shall present the strategy of this analysis,
the choice of the parameters and the observables. The results will be
presented in section 3. Our conclusions and future outlook will be
given in section 4.

\section {Strategy, the choice of parameters and observables}

In order to analyze signals for the production of squarks and gluinos
in a systematic way the large $\MGLU - \MSQ$ parameter space is
usually divided into the following three broad regions: \\
\noindent
({\bf A}) $\MSQ \gg \MGLU$,  ({\bf B}) $\MSQ \approx \MGLU$ and 
({\bf C})  $\MSQ \ll \MGLU$. 

Out of these, region (C) is not accessible to first and second
generation squarks in any grand desert scenario: even if their masses
vanish at the GUT scale, large contributions from gluino--quark loops
imply that at the weak scale $\MSQ \gsim 0.85 \MGLU$. However, some
third generation squarks might be significantly lighter than the
gluino.

Our analysis will be based on the following strategy. We shall choose
a set of GUT scale parameters {\it viz.}, $\M0$ and $\MG$, which
evolves into weak scale squark and gluino masses in regions (A) and
(B) in the conventional scenario. Moreover, $\MG$ is chosen such that
the resulting chargino mass is always consistent with the LEP bounds.
As mentioned in the Introduction, we assume that the masses of
$\tilde{b}$ and $\tilde{\tau}$ at scale $\MGUT$ are the same as those
of the corresponding first generation sfermions, i.e. we ignore the
effect of RG scaling between $\Lambda$ and $\MGUT$. Since we will
concentrate on a moderate value of $\tan \! \beta$, this should be a
good approximation for both scalar $\tau$ states as well as for
$\tilde{b}_R$. This assumption will probably over--estimate the masses
of third generation $SU(2)$ doublet squarks slightly, but we expect
this effect to be smaller than the new $D-$term contributions which
are the main focus of our study.\footnote{Note that we do not assume
the Yukawa couplings to unify. In the simplest $SO(10)$ model this
assumption would require a large value of $\tan \! \beta$, leading to
reduced masses for all third generation sfermions. Signals requiring
the presence of energetic electrons or muons will then be greatly
depleted even in the conventional scenario \cite{largetanbeta}.}

We have chosen $\tan \! \beta = 3$ and the trilinear coupling $A_0 = 0$.
The latter choice of course has no direct bearing on the
phenomenology of the first two generations, which is the focal point
of our attention. Now for $D = 0$ the magnitude of the Higgsino mass
parameter $\mu$ is fixed by the radiative electroweak symmetry
breaking condition \cite{13}. We shall work throughout with negative
$\mu$, since otherwise LEP searches require $\MGLU$ to be too large
for gluino searches at the Tevatron to be viable.

In the presence of nonuniversal Higgs masses the magnitude of $\mu$
fixed from the EW symmetry breaking condition can in principle be
quite different from that in the conventional scenario
\cite{9}. However, in order to compare our results with the
predictions of the conventional scenario, we shall work with the same
$\mu$ as in the $D = 0$ case. This does not affect the cross sections
for the production of the sparticles we are interested in, with the
sole exception of the light scalar top eigenstate $\tilde{t}_1$. Its
mass depends on $\mu$ through the off--diagonal entries of the stop
mass matrix. However, as will be discussed in more detail below, even
a rather light $\tilde{t}_1$ would not contribute very much to the
signals we are interested in. We have therefore ignored its
contribution completely; however, we have checked that the value of
$m_{\tilde{t}_1}$ that follows from our (in this case not very
reliable) assumptions satisfies the relevant search limits.

The value of $\mu$ also affects the leptonic branching ratio of the
second neutralino $\N2$ if it decays into 3--body modes. The reason is
that the $Z \LSP \N2$ coupling only proceeds through the higgsino
components of both neutralinos, and therefore scales roughly $\propto
1/\mu^2$. As a result, slepton exchange contributions to $\N2$ decay
can compete with $Z$ exchange contributions as long as slepton masses
do not greatly exceed $|\mu|$. However, we will see that our model
already allows considerable variation of this branching ratio even for
fixed $\MG$ and $\M0$; allowing $\mu$ to vary away from the value
predicted in the universal scenario would therefore not lead to new
(combinations of) signatures, although it might shift the regions in
parameter space where final states containing leptons are important.

Having chosen the parameters, it is now straightforward to compute
cross sections at the parton level in both scenarios. We shall
consider the cross section $\sigma_{n}$ corresponding to events with
$n$ leptons + jets + missing transverse energy ($E \llap {$/$}_T$)
coming from all types of squark and gluino (excluding $\tilde{t}$)
production channels. In particular $\sigma_0$ and $\sigma_2$ will turn
out to be crucial observables in our analysis. We also include the
cross section of hadronically quiet trilepton events $\sigma_{3l}$ in
our set of observables. These arise from $\CH1 - \N2$ production and
depend strongly on the slepton masses through the leptonic branching
ratio of $\N2$. In this exploratory study we do not attempt any event
simulation, nor do we apply the cuts required to extract these
signals. We should clarify that we assume the experimental definition
of the hadronic observables $\sigma_n$ to include fairly hard cuts on
the energies of the jets and/or on the missing transverse energy; this
is why we do not include contributions from the production of
electroweak gauginos here, even though they may give the biggest
contribution to the total SUSY cross section at the Tevatron. The same
hard cuts would also greatly reduce the contribution of a light
$\tilde{t}_1$ to these final states. Moreover, we only count
``primary'' electrons and muons as ``leptons''; we ignore leptons that
may result from the decay of $c-$ or $b-$ quarks or $\tau-$leptons.

In the absence of an event simulation which gives the efficiencies of
the kinematical cuts to reduce the SM background, a full appreciation
of the observability of the signals we have considered is not
possible.  Fortunately published results of such Monte Carlo
simulations allow us to at least estimate the minimal cross sections
(before cuts) that lead to detectable signals at future runs of the
Tevatron. The present bounds, from data taken during Run I, can be
gleaned from experimental searches. Both these searches and
theoretical studies of the reach of Tevatron upgrades have generally
been performed in the conventional scenario. However, we believe that
the efficiencies should not change by more than a factor of 2 when
allowing non--vanishing values of $D$, as long as we compare scenarios
with similar values of $\MG$.

Our estimates for the minimal detectable values of $\sigma_0,
\sigma_2$ and $\sigma_{3l}$ are collected in Table 1. The limits on
$\sigma_0$ and $\sigma_{3l}$ from previous Tevatron runs have been
extracted from experimental searches \cite{3,tev3l}. The future
sensitivity in the jets plus missing $E_T$ channels ($\sigma_0,
\sigma_2$) as well as in the tri--lepton channel ($\sigma_{3l}$) has
been estimated using results from refs.\cite{4} and \cite{new3l},
respectively. We also used ref.\cite{4} to estimate the sensitivity of
Run I to $\sigma_2$, since the only published experimental analysis of
the jets plus di--lepton channel \cite{cdf2l} is based on just 19
pb$^{-1}$ of data.\footnote{Both CDF and D\O\ have announced
\cite{new2l} preliminary limits from searches for final states
containing two leptons. However, the CDF search only covers the
signal from like--sign di--leptons, which according to ref.\cite{4}
has a somewhat smaller reach than the inclusive di--lepton sample,
while the D\O ~search did not include events containing muons.}

\vspace*{8mm}
\begin{center}
\begin{tabular}{|c|c|c|c|} \hline
Signal    & Run I  &  Run II& TeV33   \\ \hline \hline
$\sigma_0$& 5 &  0.9 & 0.10           \\  \hline
$\sigma_2$ & 0.6 &  0.14 & 0.03            \\  \hline
$\sigma_{3l}$& 0.25  & 0.03 & 0.009             \\  \hline
\end{tabular}
\end{center}
\vskip 5pt
{\footnotesize Table 1: The minimal detectable values of the cross
sections for jets plus missing $E_T$ final states with zero
($\sigma_0$) or two ($\sigma_2$) electrons or muons, as well as for
the relatively hadron--free three lepton final state
($\sigma_{3l}$). The second, third and fourth column show the
sensitivity from present data, from Run II (2 fb$^{-1}$ at $\sqrt{s} =
2$ TeV) and from TeV33 (25 fb$^{-1}$ at $\sqrt{s} = 2$ TeV),
respectively. All cross sections are in pb.}

\vspace*{10mm}

\section{The spectra  and the signals}

\setcounter{footnote}{0}
The sfermion masses at the GUT scale can be parametrized by \cite{9}:
\ben \label{e1} \beq
\MUL^2 &= \MUR^2 =\MSELR^2 = \M0^2 + 0.5 D \M0^2; 
\label{e1a} \\
\MDR^2 &= \MSELL^2 = \M0^2 -  1.5 D \M0^2,
\label{e1b}
\eeq \een
where the unknown parameter $D$ can be of either sign. As stated
earlier we use the same $\M0$ for the third generation, which amounts
to neglecting the running of their masses above $\MGUT$.

Requiring positivity of the sfermion mass squares at the GUT scale
imposes the restriction $-2 \leq D \leq 0.66$. Slightly stronger
constraints on $D$ arise from experimental searches. In particular,
LEP searches \cite{14} for $SU(2)$ doublet and singlet sleptons
translate into upper bounds on $|D|$ for positive and negative $D$,
respectively. For large $\M0$, the LEP bound \cite{14a} on the mass of
the lighter sbottom eigenstate $\tilde{b}_1$ gives the most stringent
bound on $|D|$ for $D<0$.

\hspace*{8mm}
\begin{center}
\begin{tabular}{|c||c|c|c|c|c|} \hline
$D$    & 0  & 0.4 &0.6   &-0.75    & -1.25     \\ \hline
$\MUL$ & 646.7 &702.4 &728.6&526.6& 428.3           \\  \hline  
$\MDL$ & 650.7 &706.0 &732.2&531.5& 434.2           \\  \hline
$\MUR$ & 644.3 &700.2&726.5 &523.7& 424.6            \\  \hline
$\MDR$ & 645.1 &436.7&279.3 &915.8& 1058.5     \\  \hline
$\MBL$ & 525.6 &592.7 &623.6&367.9& 203.5     \\  \hline
$\MSELL$ &606.1 &376.7&170.9&888.8& 1035.2     \\  \hline
$\MSELR$ & 602.6&662.1&689.8&471.5& 358.3     \\  \hline
$\MSNU$ & 601.8 &369.9&155.2&885.9& 1032.8     \\  \hline
\end{tabular}
\end{center}
\vskip 5pt 
{\footnotesize Table 2: Squark and slepton masses in GeV at the weak
scale for different values of $D$ with $\M0=600$ GeV and $\MG=105$ GeV.}

\hspace*{10mm}

We first study the signals for the choice $\M0 \approx 6 \MG$, which
corresponds to region (A) introduced in the last section. This is the
for our purposes most interesting region, since here a signal at the
Tevatron or its upgrades is not expected in the conventional
scenario. Examples for sfermion masses in this region of parameter
space are given in Table~2, for $\M0 = 600$ GeV and $\MG = 105$ GeV
and several values of $D$. The gaugino spectrum is practically
independent of $D$: $\MGLU$ = 332.1 GeV, $\MCH1$ = 95.3 GeV, $\MN2$
= 95.5 GeV, and $\MLSP$ = 46.5 GeV.

Two points are worth noting : 
\begin{itemize}
\item The masses of $SU(2)$ doublet sleptons $\LL$ and $\DR-$type
squarks can be reduced significantly for $D>0$; for $D \gsim 0.55, \
\tilde{g} \rightarrow \DR d$ 2--body decays become possible.
\item Due to the smaller coefficient of $D$ in eq.(\ref{e1a}) as
compared to (\ref{e1b}), the reduction of the masses of $\UL, \DL$ and
$\UR$ squarks as well as $\LR$ sleptons for $D<0$ is at first less
significant. However, for large negative $D$ ($D \leq -0.8$) the $\BL$
becomes lighter than the gluino; in fact, as mentioned in the previous
section, the lower bound on $D$ comes from the experimental bound on
$m_{\tilde{b}_1}$ in this scenario.\footnote{A recent study
\cite{baer10} of the sparticle spectrum predicted by the minimal
$SO(10)$ SUSY GUT also emphasized the reduction of $m_{\tilde{b}_1}$
relative to the other squark masses. However, as mentioned in Sec.~2,
minimal $SO(10)$ requires large values of $\tan \!
\beta$, leading to a large bottom Yukawa coupling. Successful
electroweak symmetry breaking then requires $D > 0$. As a result, the
light $\tilde{b}_1$ found in ref.\cite{baer10} is mostly an $SU(2)$
singlet, while our more general scenario allows $D<0$ and a light $\BL$.}
\end{itemize}
 
These changes of the scalar spectrum lead to intriguing features of
gaugino branching ratios. In the electroweak gaugino sector, BR$(\N2
\rightarrow \LSP l \bar{l}$) increases significantly for $D>0$, due to
the reduction of $m_{\tilde{l}_L}$.  For example, we have BR$(\N2
\rightarrow \LSP l \bar{l}$) = 0.02, 0.07 and 0.17 (for each lepton
flavor) for $D=$ 0, 0.4 and 0.6, respectively. As shown in Fig.~1,
for $D \lsim 0.5$ this results in an appreciable enhancement of the
dilepton signal ($\sigma_2$) compared to the conventional
scenario. The clean trilepton signal ($\sigma_{3l}$) grows
monotonically with increasing $D$. 

Reducing $D$ below zero at first leads to little change of BR$(\N2
\rightarrow \LSP l \bar{l}$). Note that in the conventional scenario
an increase of $\M0$ beyond 600 GeV results in an increase of this
branching ratio, due to reduced destructive interference between $Z$
and slepton exchange contributions. In our case reducing $D$ below
zero leads to lighter $SU(2)$ singlet sleptons, but heavier doublet
sleptons; the two effects appear to largely cancel in the leptonic
branching ratio of $\N2$. However, for $D \lsim -1.2$ the reduced
$\BL$ mass leads to a rapid increase of $BR(\N2 \rightarrow \LSP b
\bar{b})$, and a corresponding reduction of the leptonic branching
ratio, as well as of $\sigma_{3l}$.

The chargino decays do not exhibit any remarkable feature over the
entire region of $D$ considered above, since they are dominated by
virtual $W$ exchange contributions.

The gluino branching ratios for $D =$ 0 and 0.4 are:\footnote{We
ignore a small contribution from $\GLU \rightarrow \CH1 t b$. For our
choice of gluino mass the corresponding partial width is suppressed by
phase space. Moreover, its exact value depends sensitively on the
masses and mixing angle of the stop squarks, which in turn depend on
details of the GUT model, as discussed earlier.}

\noindent
$BR(\GLU \rightarrow \CH1 q \bar{q^{\prime}}) = 0.47,\ 0.37 \ \ 
(qq^{\prime} = ud, \ cs)$,\\
$BR(\GLU \rightarrow \LSP q \bar{q}) = 0.14,\ 0.35 \ \ 
(q = u,\ d,\ c,\ s,\ b)$, \\
$BR(\GLU \rightarrow \N2 q \bar{q}) = 0.39,\ 0.28 \ \  
(q = u,\ d,\ c,\ s,\ b)$.

Note that the branching ratio for $\GLU \rightarrow \LSP q \bar{q}$
more than doubles when going from $D=0$ to $D = 0.4$. This is a
consequence of the reduced $\DR$ masses; recall that $SU(2)$ singlet
squarks couple only very weakly to $\N2$ and $\CH1$, which are
predominantly $SU(2)$ gauginos. This leads to an increase of
$\sigma_0$ (see Fig.~1), a hardening of the missing $E_T$ spectrum,
and an increase of the fraction of gluino pair events containing
$b-$jets.

These changes become even more dramatic for $D \geq 0.55$. Now
$\DR-$type squarks become lighter than the gluino, so that all
gluinos decay to $\DR$s with $BR = 1$. Almost all the $\DR$'s in turn
decay directly into the LSP. This suppresses $\sigma_2$ severely
compared to the conventional case. In addition, $\GLU \DR$ production
as well as the pair production of $\DR-$type squarks lead to a further
increase of $\sigma_0$, by up to a factor of 2. Recall that the clean
trilepton signal ($\sigma_{3l}$) is also enhanced relative to the
universal case. This combination of significant $\sigma_0$ and
$\sigma_{3l}$ but nearly vanishing $\sigma_2$ cannot be accommodated
in the conventional scenario for any choice of the parameters.

As mentioned above, taking $D < 0$ decreases the masses of $SU(2)$
doublet as well as $\UR-$type squarks. This leads to increased
destructive interference between $s-$ and $t-$channel contributions to
$\GLU$ pair production; as a result, the corresponding cross section
drops by about a factor 1.6 when $D$ is decreased from 0 to $-0.9$. We
also saw that the decrease in squark mass is especially pronounced for
$\BL$, whose mass is also reduced by RG--running from the GUT to the
weak scale. Reducing $D$ below 0 therefore also increases the fraction
of $\GLU$ decays that produce $b-$jets; we saw a similar effect for
$D>0$ due to the reduction of $m_{\tilde{b}_R}$. However, while this
fraction saturates at $1/3$ for $D > 0$, where all three $\DR-$type
squarks become light, it reaches 100\% for $D <
-0.9$, where $\GLU \rightarrow \BL b$ is the gluino's only 2--body
decay mode. Nearly all $\BL$ in turn decay into $\N2 b$. Since
leptonic $\N2$ decays always lead to lepton pairs with opposite
charges, the fraction of like--sign di--lepton pairs in $\sigma_2$
becomes very small here.

For $D \lsim -1$, $\BL$ pair production begins to contribute
significantly to $\sigma_0$ and $\sigma_2$. For example, for $D = -1.3
\ (-1.4)$ we have $m_{\tilde{b}_L} = 174 \ (115)$ GeV, giving
$\sigma(\BL \BL^*) \simeq 1.0 \ (8.5)$ pb. Due to the presence of (at
least) two hard $b-$jets in the event, a dedicated sbottom pair search
is expected to be sensitive to much smaller cross sections than more
generic SUSY searches. A recent analysis \cite{pedro} concludes that
data from from Run II (TeV33) should allow to search for $\BL$ with
mass up to at least 185 (205) GeV even if all $\N2$ decay into light
quarks; in the case at hand this would cover the range $D \lsim
-1.25$. Scenarios with $D \lsim -1.4$ can perhaps even be probed with
existing data from Run I.\footnote{A search by the D\O ~collaboration
\cite{15} was able to exclude $m_{\tilde{b}_1} \leq 115$ GeV if
$BR({\tilde{b}_1} \rightarrow \LSP b) \simeq 100\%$ and $\MLSP \lsim 20$
GeV; neither of these conditions is satisfied in our case. However,
during Run I CDF had significantly better $b-$tagging capabilities
than the D\O ~detector.} Recall also that for large negative $D$,
$BR(\N2 \rightarrow \LSP b \bar{b})$ is enhanced, leading to events
with up to 6 $b-$jets! In the conventional scenario one would expect
$\sigma_0$ and $\sigma_2$ to be dominated by sbottom pair production
only at very large values of $\tan \! \beta$; and even there the
$\tilde{b}_1 - \GLU$ mass splitting is less than for large negative
$D$ in the non--universal case. Moreover, in the conventional scenario
one also expects final states to be rich in $\tau-$leptons
\cite{largetanbeta}; this need not be the case in the non--universal
scenario. 

It is clear from Table~1 and Figure~1 that if $\MSQ \gg \MGLU$, the
latter being tightly constrained by LEP bound on $\MCH1$, the
conventional scenario offers very little hope for a discovery in any
of the three channels up to the TeV33 phase. In contrast, the
non--universal scenario allows for cross sections in the striking
range of Run II experiments, from sbottom pair production for large
negative $D$, and in the tri--lepton channel at large positive
$D$. Unfortunately observation of a signal in the tri--lepton channel
by itself would not be sufficient to discriminate between the two
scenarios. In the case at hand this distinction should be possible
using TeV33 data, since there $\sigma_0$ and $\sigma_{3l}$ become
measurable over the entire range of $D$ shown in Fig.~1. As noted
earlier, the combination of large $\sigma_0$ and $\sigma_{3l}$ with
small $\sigma_2$ would exclude the conventional scenario. Additional
clues can be obtained by correlating kinematical distributions, which
are sensitive to sparticle masses, with measurements of the total cross
sections. 

The nonuniversal model allows for an observable trilepton signal even
for higher $\MGLU$, albeit for much lighter sleptons ($D \gsim
0.6$). For example, with $D = 0.6$, $\MG$ = 130 GeV ($\MGLU$ = 420
GeV), $\MLL$ = 190 GeV, $\S3l$ = 0.09 pb should be easily detectable
already at Run II. The corresponding $\sio$ is, however, unobservable
even at TeV33.

We next consider region (B) of the parameter space. We shall work
with $\MG$ = 105 GeV and $\M0 = 300$ or 200 GeV.
This is because even the phenomenology of the conventional scenario
depends sensitively on $\M0$.
Additional comments will be made on higher $\MGLU$ as and when appropriate.
For $\M0$ = 300 GeV, the following mass patterns are obtained:

\vspace*{8mm}
\begin{center}
\begin{tabular}{|c||c|c|c|c|c|} \hline
$D$    & 0  & 0.5 &-1.00   &-1.75        \\ \hline
$\MUL$ & 392.6 &421.5 &327.4&268.3     \\  \hline
$\MDR$ & 389.9 &285.6 &541.2&631.3     \\  \hline
$\MBL$ & 339.9 &372.3 &261.1&181.6     \\  \hline
$\MSELL$ &312.2 &164.3&488.2&586.5     \\  \hline 
$\MSNU$ & 303.9 & 147.9 & 482.9 & 582.1 \\ \hline
$\MSELR$ &305.3 &341.6&215.0&105.0     \\  \hline
\end{tabular}
\end{center}
\vskip 5pt 
{\footnotesize Table 3: Squark and slepton masses (in GeV) at the weak
scale for different values of $D$ with $\M0=300$ GeV and $\MG=105$
GeV.}

\vspace*{10mm}
\noindent
The spectrum for the gaugino sector is: $\MGLU=320$ GeV (the
difference from the previous case arises due to radiative corrections
to the gluino pole mass from squark--quark loops), $\MCH1$ = 95.16 GeV,
$\MN2$ = 95.06 GeV, and $\MLSP$ = 46.71 GeV.

Points to be  noted are : 
\begin{itemize}
\item In the conventional scenario the squarks are somewhat heavier
than the $\GLU$.
\item Increasing $D$ from zero again reduces $\MDR$ and $\MLL$;
in particular, for $D \geq 0.5, \ \MDR < \MGLU$ for all three
generation of $\DR-$type squarks; and
\item For large negative $D$ ($D \leq -1$) the $\BL$ again becomes
lighter than the gluino, but the theoretical lower bound on its mass
is higher than in the previous case, because of the LEP bounds on
$\MSELR$. 
\end{itemize}

In contrast to the previous case, all possible squark and gluino
production channels can contribute to $\sigma_0$ and $\sigma_2$. As
shown in Fig.~2, the size of these cross sections will be typically
larger than the conventional case even for moderate magnitudes of $D$
irrespective of its sign. $\sigma_0$ still stays below the current
bound of about 5 pb over the entire range of $D$; however, the region
with $D \lsim -1$ should be accessible in this channel at Run II, and
the entire range of $D$ can be probed at TeV33.

Unlike in the scenario with large $\M0$ we now observe a significant
increase of the leptonic branching ratio of $\N2$ for large negative
$D$.  Note that now slepton exchange contributions dominate the
leptonic partial width of $\N2$ even in the conventional
scenario. Decreasing $D$ away from zero therefore at first actually
reduces the leptonic branching ratio of $\N2$, due to the increase of
$\MSELL$; note that for roughly equal mass the exchange of $SU(2)$
doublet sleptons contributes more, due to their stronger coupling to
$\N2$. However, for $D \lsim -1$ the reduction of $\MSELR$
over--compensates the increase of $\MSELL$, leading to an increase of
$\sigma_{3l}$; in fact, the region around $D = -1.75$ might already be
excluded by existing searches \cite{tev3l} for tri--lepton events. For
even more negative values of $D$, the decay $\CH1 \rightarrow
\tilde{\tau}_1 + \nu_\tau$ opens up. Since we do not include secondary
leptons from $\tau$ decays when computing $\sigma_{3l}$, this cross
section essentially vanishes once that decay mode dominates. In
reality a measurable tri--lepton signal may survive, if softer cuts
on the leptons are employed \cite{new3l}; in addition, one may look
for events with two leptons and a $\tau-$jet \cite{largetanbeta}. Note
that unlike $\CH1, \ \N2$ also couples to $\tilde{e}_R$ and
$\tilde{\mu}_R$; the leptonic branching ratio of $\N2$ therefore
remains large all the way out to the LEP--imposed lower bound on $D$.

Together with the reduction of $\MBL$ this can lead to promising
signals from events with at least one hard $b-$jet and two
opposite--charge, same--flavor leptons; this might extend the reach of
Run II to smaller values of $|D|$ than those accessible to sbottom
searches based on final states without leptons \cite{pedro}. Even
these kinds of searches should be able to probe $D \lsim -1.75 \
(-1.6)$ at Run II (TeV33). Moreover, even a moderately negative $D$
leads to a significant increase of $\sigma_2$. Recall that in this
region all $SU(2)$ doublet squarks, as well as $\UR-$type squarks, are
somewhat lighter than in the conventional scenario (for the same value
of $\MG$ and $\M0$), leading to a significant increase of the total
cross section for the production of strongly interacting
sparticles. This over--compensates the modest decrease of the
leptonic branching ratio of $\N2$ for moderately negative
$D$. Nevertheless this region of parameter space can be probed in the
inclusive di--lepton channel only at TeV33. 

Increasing $D$ from zero leads to quickly increasing leptonic
branching ratio of $\N2$, and a corresponding rise of $\sigma_{3l}$;
as for negative $D$, the region just below the largest value of $|D|$
allowed by slepton searches at LEP is again at best marginally
consistent with current search limits for tri--lepton final
states. For the largest LEP--allowed value of $D$, the decays $\N2
\rightarrow \tilde{\nu} \nu$ open up. The sneutrino then decays into
the invisible channel $\tilde{\nu} \rightarrow \LSP \nu$ (the
so--called VLSP scenario \cite{18}), leading to a vanishing
$\sigma_{3l}$. Apart from the most extreme regions, the entire range of
$D$ can easily be covered in the tri--lepton mode at TeV33, and much
of it should already be accessible to Run II experiments. Finally,
sizably positive $D$ again suppresses $\sigma_2$, first due to the
dominance of $\GLU \rightarrow \DR d$ decays (for $D \geq 0.35$), and,
for the largest allowed value of $D$, due to $\N2 \rightarrow
\tilde{\nu} \nu$ decays.

For $\M0 = 200$ we obtain the following sfermion spectra: \\

\hspace*{8mm}
\begin{center}
\begin{tabular}{|c||c|c|c|c|c|} \hline
$D$    & 0  & 0.4 &0.6   &-0.75& -1.75     \\ \hline
$\MUL$ &326.6 &339.1 &345.2&301.7&264.4           \\  \hline
$\MDR$ &323.4 &282.0&258.8&389.3&462.7    \\  \hline
$\MBL$ &293.6 &307.5 &314.2&265.6&222.9    \\  \hline
$\MSELL$ &218.0 &149.9&99.7&307.4&396.3     \\  \hline
$\MSELR$ &207.8 &227.0&236.1&166.0& 81.7     \\  \hline
$\MSNU$ &205.9 &131.7&69.5&298.9& 389.8     \\  \hline
\end{tabular}
\end{center}
\vskip 5pt 
{\footnotesize Table 4: Squark and slepton masses (in GeV) at the weak
scale for different values of $D$, for $\M0=200$ GeV and $\MG=105$ GeV.}

\vspace*{10mm} 

The gaugino masses are nearly the same as for $\M0=300$ GeV. The
behavior of our three cross sections, shown in Fig.~3, is
qualitatively similar to the previous case. However, due to the
reduced importance of the $D-$terms as compared to the universal
contribution from gluino--quark loops, even in the non--universal
scenario the mass splitting between different first or second
generation squarks now amounts to at most 50\%. As a consequence the
changes of the three cross sections $\sigma_0, \ \sigma_2$ and
$\sigma_{3l}$ with varying $D$ are somewhat smaller than for larger
values of $\M0$. In particular, $\sigma_0$ now increases by less than
a factor of 3 as $D$ is decreased from zero to its most negative
allowed value. On the other hand, owing to the reduction of the
overall squark mass scale, the entire range of $D$ can now be probed
in the jets plus missing $E_T$ channel at Run II of the Tevatron.

Somewhat larger variations can occur for the leptonic cross
sections. Since sleptons are still significantly heavier than $\N2$
and $\CH1$ in the conventional scenario ($D=0$), the leptonic
branching ratios can change considerably when $D$ is allowed to
take values close to the present upper or lower bounds. Due to the
reduction of the slepton masses compared to the previous cases, these
cross sections are also generally larger than for scenarios with
larger $\M0$. In particular, the tri--lepton signal should be
detectable over the entire range of $D$ at Run II. The availability of
at least two distinct SUSY signals should make it easier to
distinguish between the universal and non--universal scenarios; on the
other hand, the expected differences in the squark spectrum are
smaller than before, as already noted.

Yet smaller values of $\M0$ continue the trend of going from 300 to
200 GeV. The maximal possible mass splitting between squarks keeps
decreasing even in the non--universal scenario. Significant
differences in the slepton spectrum remain possible; as long as
sleptons are heavier than $SU(2)$ gauginos in the conventional
scenario, these changes of slepton masses affect the leptonic
branching ratios of electroweak gauginos. However, for $\M0 \leq 85$
GeV, these gauginos can decay into on--shell sleptons even in the
universal scenario; their leptonic branching ratios are then much less
sensitive to $D$. Moreover, LEP searches severely limit the maximal
allowed value of $D$ for such small values of $\M0$.

\section{Summary and conclusions}

In this paper we have studied the impact of $SO(10)$ $D-$term
contributions to scalar masses on the search for supersymmetric
particles at the Tevatron collider. These contributions have opposite
signs for sfermions residing in the {\bf 5}$^*$ representation of
$SU(5)$ and those residing in the {\bf 10} representation. Any
non--zero value of $D$ will therefore reduce the masses of some
sfermions, while making others heavier. This usually results in an
increase of the total cross section for sparticle production at hadron
colliders. Heavy sparticles simply decouple, so that further increases
of their masses have little impact on any observables; however, the
cross section for the production of the lighter sfermions keeps
increasing as their masses are reduced. The only exception to this
rule occurs for large $m_0$ and moderately negative $D$, where the
reduction of most squark masses actually reduces the gluino pair cross
section through interference effects. However, we found that this can
reduce the total cross section for the production of strongly
interacting sparticles by no more than a factor of 2; in contrast, for
extreme (but allowed!) values of $D$ this cross section can be
enhanced by more than an order of magnitude compared to models with
degenerate squarks.

Similarly, non--degeneracy in the slepton sector tends to increase
leptonic branching ratios of electroweak gauginos, in particular those
of $\N2$, leading to an increase of the relatively hadron--free
tri--lepton cross section from $\N2 \CH1$ production. For large
negative $D$ this cross section can be sufficiently large to yield
unambiguous signals at the next run of the Tevatron collider even if
the {\em average} sfermion mass lies well above the gluino mass, which
in turn lies well above its lower limit derived from chargino searches
at LEP (assuming gaugino mass unification). However, if $D$ is very
close to its LEP--imposed upper or lower limit, the tri--lepton signal
is greatly depleted, since there decays $\N2 \rightarrow \LSP
\tilde{\nu} \nu$ and $\CH1 \rightarrow \tilde{\tau}_1 \nu_\tau$,
respectively, become possible.

One interesting feature of the non--universal scenario is that one can
find combinations of parameters where the tri--lepton and multi--jet
plus missing $E_T$ signals are both large, but the jets plus
di--lepton plus missing $E_T$ signal is very small. This last signal
results from cascade decays of the gluino and $SU(2)$ doublet squarks,
which are suppressed for $D > 0$, where $\DR-$type squarks can be the
only light squarks. Such a combination of signals cannot occur if
all sfermions are degenerate at the GUT scale.

Models with non--vanishing $D$ also introduce new signals, which do
not hold much promise if squarks are approximately degenerate. These
signals originate from the copious production of light sbottom
squarks. In models with small or moderate $\tan \! \beta$ studied
here, light $\BL$ squarks can occur for negative $D$. Since
$\BL$ dominantly decays into $\N2 b$, $\BL$ pair production (as well
as gluino production followed by $\GLU \rightarrow \BL b$ decays)
gives rise to several signals. Since for large negative $D$ the
branching ratio for $\N2 \rightarrow \LSP b \bar{b}$ decays can be
enhanced, searches for final states containing 4, or even 6, $b-$jets
may offer the best reach. For somewhat smaller $|D|$ or smaller $m_0$,
searches for final states containing two hard $b-$jets and two
opposite--charged, same--flavor leptons might hold even more
promise. This reinforces the conclusions of ref.\cite{baer10}
regarding the importance of SUSY signals containing hard $b-$jets.

The maximal possible mass splitting in the squark sector becomes
smaller for smaller ratios of the GUT scale parameters $\M0/\MG$, due
to the increased importance of flavor--universal contributions to
squark masses from gluino--quark loops. For $\M0 < 2 \MG$ the main
impact of sfermion non--universality on SUSY signals at the Tevatron
comes from changes of the leptonic branching ratios of electroweak
gauginos, which remain sensitive to $D$ as long as $\M0 \gsim
\MG$. For even smaller values of $\M0/\MG$ evidence for
non--universality in the sfermion sector can probably best be found in
studies of slepton pair production. However, these may have to be
postponed until a lepton supercollider has been built.

In summary, in $SO(10)$ models with sizable $D-$term contributions to
scalar masses, SUSY searches at future collider runs of the Tevatron
hold even more promise than in the conventional mSUGRA model. New
regions of the $(\M0, \MG)$ plane become accessible to ``standard''
searches involving jets and/or leptons plus missing transverse
momentum, and searches for new final states containing hard $b-$jets
may increase the reach even further. Under favorable circumstances the
pattern of observed signals will allow to discriminate between
universal and non--universal scenarios. This once again underscores
the importance of searching for supersymmetry in as many different
channels as possible.

\subsection*{Acknowledgements} 

We wish to thank the organizers of the Fifth Workshop on High Energy
Physics Phenomenology (WHEPP5), Pune, India (1998), where this work
was begun.  Amitava Datta and Aseshkrishna Datta acknowledge partial
financial support from a DST, India, Project, No. SP/S2/k01/97. MD was
supported by FAPESP (Brazil). Amitava Datta and D. P. Roy
acknowledges financial support from a BRNS Project, No. 37/4/97-G --
R\&D II/474.

\newpage

\begin{figure}[htb]
\vspace*{-1.in}
\centerline{
\epsfig{file=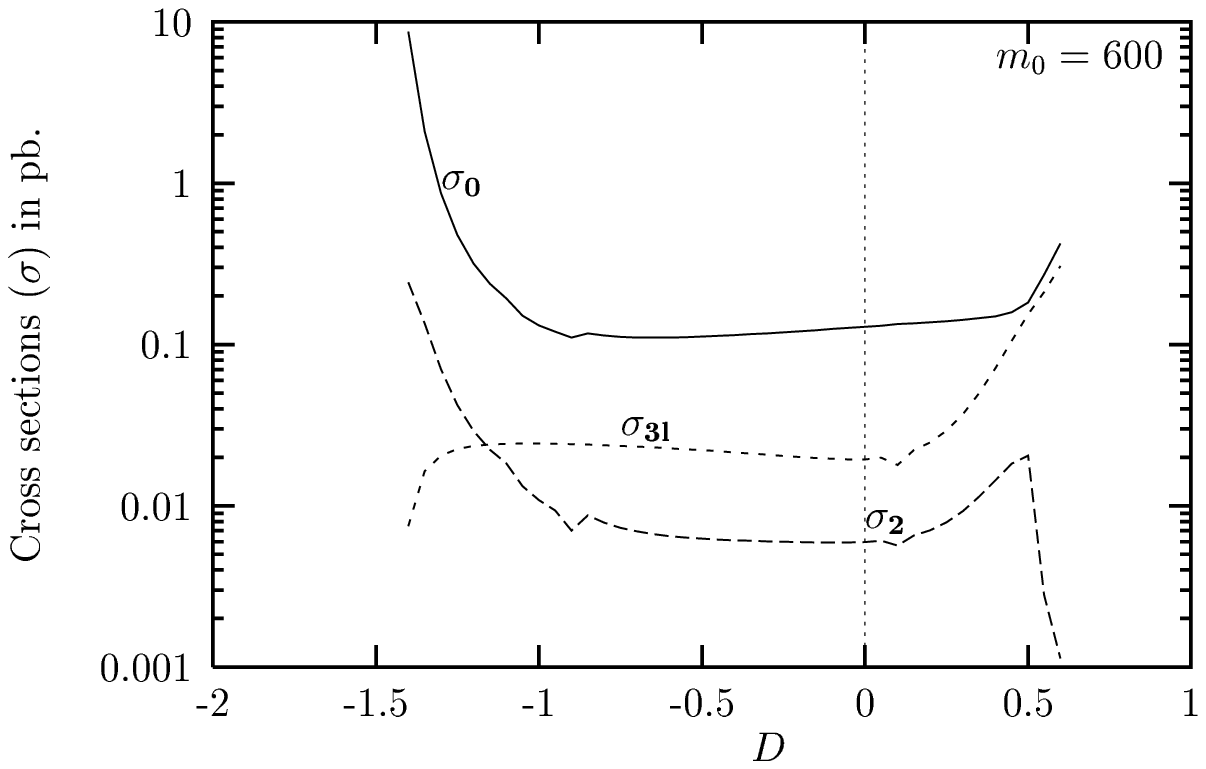, height=20cm,width=15cm}
}
\vspace{-4.5in}
\caption
{Cross sections (in pb, without any cut) for different final states at
the Tevatron collider with $\sqrt{s}=2$ TeV as a function of $D$. The
other relevant parameters are $\MG=105$ GeV, $\M0=600$ GeV and $\tan
\! \beta=3$, while $\mu$ is fixed by imposing the radiative electroweak
symmetry breaking condition for $D=0$.}
\end{figure}

\vskip 40pt
\begin{figure}[htb]
\vspace*{-1.in}
\centerline{
\epsfig{file=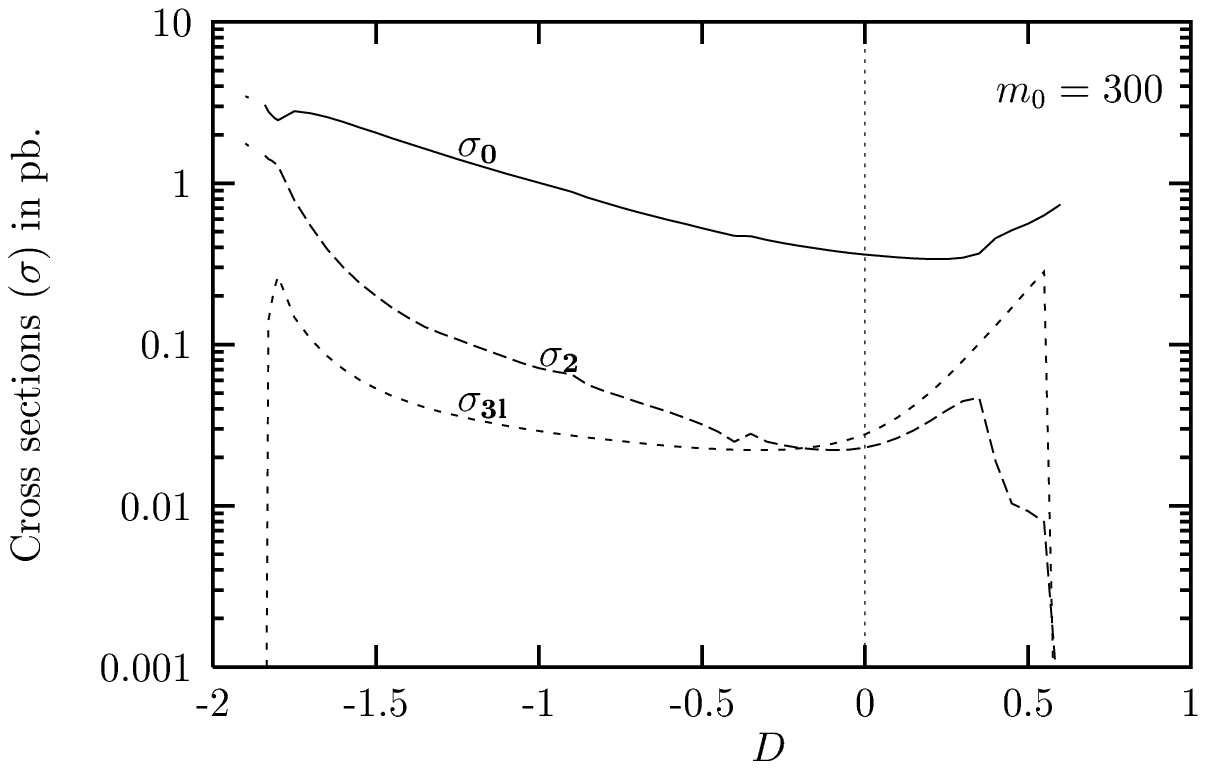, height=20cm,width=15cm}
}
\vspace{-4in}
\caption{As in Fig.~1, but with $\M0=300$ GeV.}
\end{figure}

\begin{figure}[htb]
\vspace*{-2.00in}
\centerline{
\epsfig{file=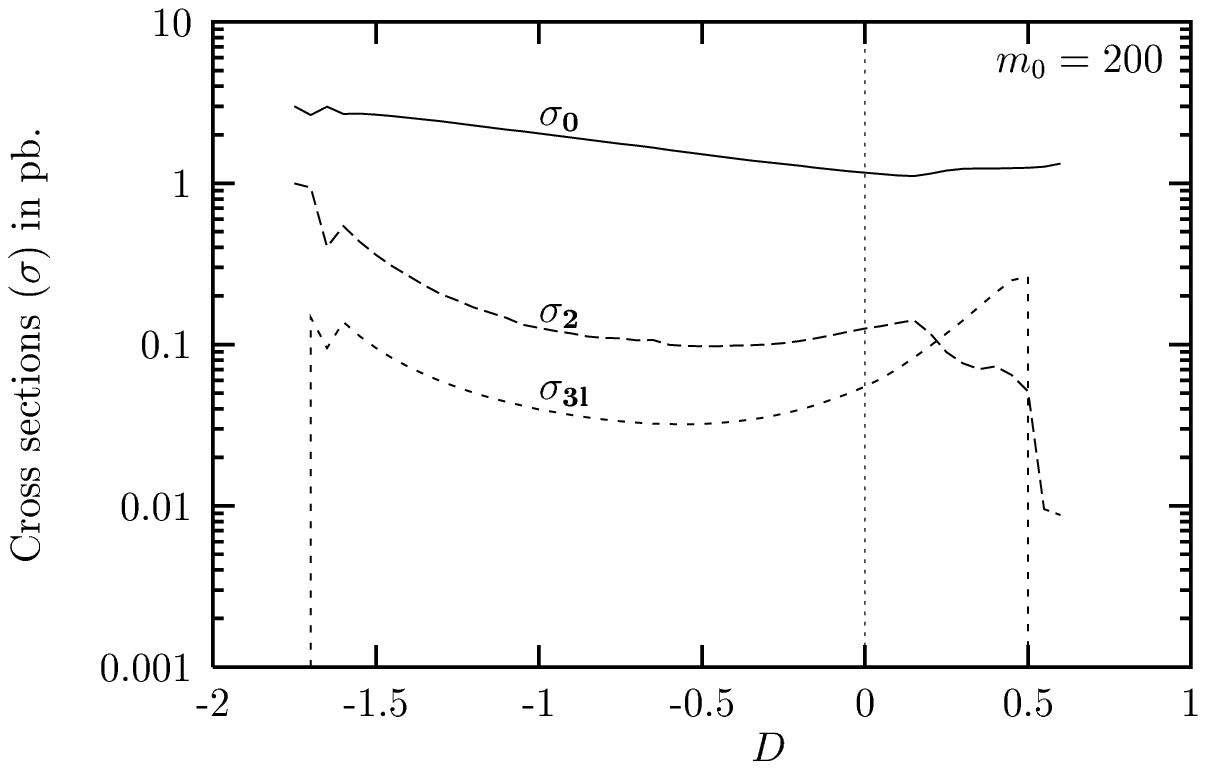, height=20cm,width=15cm}
}
\vspace{-4in}
\caption{As in Fig.~1, but with $\M0=200$ GeV.}
\end{figure}

\end{document}